\documentclass{pasj00}
\usepackage{bm}

\newcommand{\argmin}{\mathop{\rm arg~min}\limits}

\begin{document}
\SetRunningHead{M. Uemura et al.}{Doppler Tomography by Total
  Variation Minimization}

\title{Doppler Tomography by Total Variation Minimization}

\author{Makoto \textsc{Uemura}} %
\affil{Hiroshima Astrophysical Science Center, Hiroshima University,
  Kagamiyama 1-3-1, Higashi-Hiroshima, Hiroshima 739-8526}
\email{uemuram@hiroshima-u.ac.jp}

\author{Taichi \textsc{Kato}}
\affil{Department of Astronomy, Kyoto University,
  Kitashirakawa-Oiwake-cho, Sakyo-ku, Kyoto 606-8502}

\author{Daisaku \textsc{Nogami}}
\affil{Kwasan and Hida Observatories, Kyoto University, Yamashina-ku,
  Kyoto 607-8471}

\and

\author{Ronald \textsc{Mennickent}}
\affil{Universidad de Concepci\'{o}n, Departamento de Astronom\'{i}a, Casilla
  160-C, Concepci\'{o}n, Chile}


%


\maketitle

\begin{abstract}
We have developed a new model of the Doppler tomography using total
variation minimization (DTTVM). This method can reconstruct localized
and non-axisymmetric profiles having sharp edges in the Doppler
map. This characteristic is emphasized in the case that the number of
the input data is small. We apply this model to real data of the dwarf
novae, WZ~Sge in superoutburst and TU~Men in quiescence. We
confirmed that DTTVM can reproduce the observed spectra with a high
precision. Compared with the models based on the maximum entropy
method, DTTVM provides the Doppler maps that little depend on the
hyper-parameter and on the presence of the absorption core. We also
introduce a cross-validation method to estimate reasonable values of
a hyperparameter in the model by the data itself. 
\end{abstract}

\section{Introduction} 

Doppler tomography (DT) is a widely-used method in the field of binary
systems (\cite{mar88mapping}). It reconstructs an emission-line
intensity map in the velocity space (Doppler map) from the temporal
variation of the line profile. This method is valid in the case
that the line forming region co-rotates with the orbital motion. Such 
a condition is indeed a good approximation for real data of the
emission line of the accretion disk in cataclysmic variables (CVs)
(\cite{sma81diskemission}). A number of studies using DT have
revealed characteristic features of accretion disks in CVs: spiral
patterns, stream-impact region, and also the emission from the
secondary star (e.g. \cite{mar90ugem, ste97ippeg, sch97huaqr}). 

DT is an ill-posed inverse problem in most cases. A filtered
back-projection method provides a way to solve the problem. It masks
high frequency components in the Doppler maps in order to uniquely
determine the solution. The other way is to introduce regularization
terms in addition to the likelihood term. The maximum entropy method
(MEM) has been used for DT. The DT with MEM determines the solution by
maximizing the information entropy over the default image. One can
choose an appropriate default image, for example, axi-symmetric or
Gaussian blur image. The comparison between the back-projection method
and MEM is reported in \citet{mar01DopplerTomography}. 

In general, those DT methods tend to smear out localized and
sharp-edge profiles in the Doppler map because such profiles have high
frequency components which make the entropy low. On the other hand,
they are expected in accretion physics, such as spiral
shocks and hot spots. It has been reported that there are significant
residuals between the observed and model spectra that are obtained 
from DT (e.g. \cite{tap03dopmap}). It is, however, unclear whether the 
large residuals are caused by the emission components which violate the
basic assumption that the emission is isotropic and optically
thin. Alternatively, it may be caused by regularization terms which
are ill-suited for the real disk. 

In this paper, we propose a new DT by total variation
minimization (TVM). TVM has received attention in the field of image
reconstruction because it can reconstruct sharp-edge features by
making the image sparse in its gradient domain. In section~2, our model of
DT by TVM (DTTVM) is described. In section~3, several experiments of
DTTVM are shown for both artificial and real data. In section~4, we
shortly discuss the treatment of absorption components in our model,
and summarize our findings in section~5. 

\section{Model}

DT is a method to reconstruct the Doppler map, i.e., the intensity
map in the velocity space $(v_x,v_y)$. The coordinate system in our
model is the same as previous models of DT (\cite{mar88mapping}): The
radial velocity $v_{\rm R}$ of a point $(v_x,v_y)$ in the map is
represented as: 
\begin{eqnarray}
v_{\rm R} = -v_x \cos \phi +v_y \sin \phi,
\end{eqnarray}
where $\phi$ is the orbital phase that can be calculated from the
observation time and the ephemeris of the binary system. Suppose that
there are $M$ data given by $\bm{d}=\{d_0,d_1,\cdots,d_M\}$. The $i$-th
element of $\bm{d}$, $d_i$, is the flux at a radial velocity $v_{{\rm
    R},i}$ and phase $\phi_i$. Also suppose $n\times n = N$ points in
the Doppler map given by
$\bm{s}=\{s_{0,0},s_{0,1},\cdots,s_{n,n}\}$. The $j$-th element in the map has
the coordinate of $(v_{x,j},v_{y,j})$ and intensity of $s_j$. The
radial velocity at $\phi_i$ and $(v_{x,j},v_{y,j})$ can be calculated
from equation~(1) and represented as $v_{{\rm R},ij}$.

The goal of DT is to estimate $\hat{\bm{s}}$ from observations
according to 
\begin{eqnarray}
\hat{\bm{s}} = \argmin_{\bm{s}} \left\{ \frac{1}{M} \| \bm{d}-A\bm{s}
\|^2_2 + \frac{\lambda}{N} \Phi(\bm{x}) \right\}. 
\end{eqnarray}
The first term in the right side is the least-square term, and the
second term is the regularization term which consists of a
hyperparameter $\lambda$ and a function $\Phi$. In equation~(2),
$A$ is a $M\times N$ matrix whose element depends on the phase, radial 
velocity, and instrumental response. For the response function, we
assumed a Gaussian profile with a variance $\sigma^2$. An element of
$A$, $a_{ij}$, is then given by 
\begin{eqnarray}
a_{ij} = \frac{1}{\sqrt{2\pi\sigma^2}} \exp 
\left\{ -\frac{(v_{{\rm R},i}-v_{{\rm R},ij})^2}{2\sigma^2}  \right\}.
\end{eqnarray}
Here, $\sigma$ means a width of the instrument line broadening. The
sampling interval of spectra is usually designed to be smaller than the
broadening width in astronomical spectrographs. Hence, $\sigma$
corresponds to a spectral resolution in most cases of actual observations.

TVM provides a method to solve ill-posed problems. The isotropic total
variation of the map, $\Phi_{\rm TVM,iso}$, is defined as 
\begin{eqnarray}
\Phi_{\rm TVM,iso}(\bm{s}) = \sum_{i,j} \sqrt{(s_{i+1,j}-s_{i,j})^2+(s_{i,j+1}-s_{i,j})^2}. 
\end{eqnarray}
The anisotropic total variation is
\begin{eqnarray}
\Phi_{\rm TVM,aniso}(\bm{s}) = \sum_{i,j} |s_{i+1,j}-s_{i,j}|+|s_{i,j+1}-s_{i,j}|.
\end{eqnarray}
$\Phi_{\rm TVM,iso}$ is used as $\Phi$ in equation~(2) in this paper.
We confirmed that the model with $\Phi_{\rm TVM,aniso}$ generates
almost the same results as that with $\Phi_{\rm TVM,iso}$. 

We use the TwIST algorithm to estimate $\hat{\bm{s}}$ in equation~(2)  
(\cite{twist1,twist2}).\footnote{$\langle$http://www.lx.it.pt/\~bioucas/TwIST/TwIST.htm$\rangle$} 
This is a kind of Iterative Shrinkage/Thresholding (IST) algorithm, or
sometimes called as the proximal gradient method. IST is an algorithm
for minimization problems of a function that is a sum of a
differentiable convex function and non-differentiable, but simple-form
convex function. The least-square and TVM terms correspond to those
two functions in the present case. IST solves the problem by a
combination of the proximal point method and gradient method: One step
consists of a gradient step for the least-square term and a proximal
point correction step for the TVM term. The TwIST
algorithm is a two-step iterative version of IST, enabling a faster
convergence than IST. In the present case, the iteration algorithm of
TwIST is, 
\begin{eqnarray}
\bm{s}_{t+1} &=& (1-\alpha)\bm{s}_{t-1} + (\alpha-\beta)\bm{s}_t \\ \nonumber
&&+\beta\Psi_\lambda \left(\bm{s_t}+A^T(\bm{d}-A\bm{s_t})\right), 
\end{eqnarray}
where $\Psi_\lambda$ is the proximal operator. For TVM, there is no
closed form of $\Psi_\lambda$, and the iterative method proposed in
\citet{cha04tvm} is used. In the present case, we used $\alpha=1.44$
and $\beta=0.92$.The intensity
$s_{i,j}$ in the Doppler map can be negative in the present model.

The calculation code of DTTVM is available at our web
site.\footnote{$\langle$http://home.hiroshima-u.ac.jp/uemuram/dttvm/$\rangle$}
As can be seen in equation~(2), the present model assumes that the
noise follows a symmetric probability distribution, like a Gaussian
distribution, and that all observations are yielded with the same
variance. The least-square term in equation~(2) should be replaced by
the least-square term weighted with each observation error if the data
quality varies widely. 

\section{Results}

\subsection{Experiments using artificial data}

\begin{figure*}
  \begin{center}
    \FigureFile(170mm,170mm){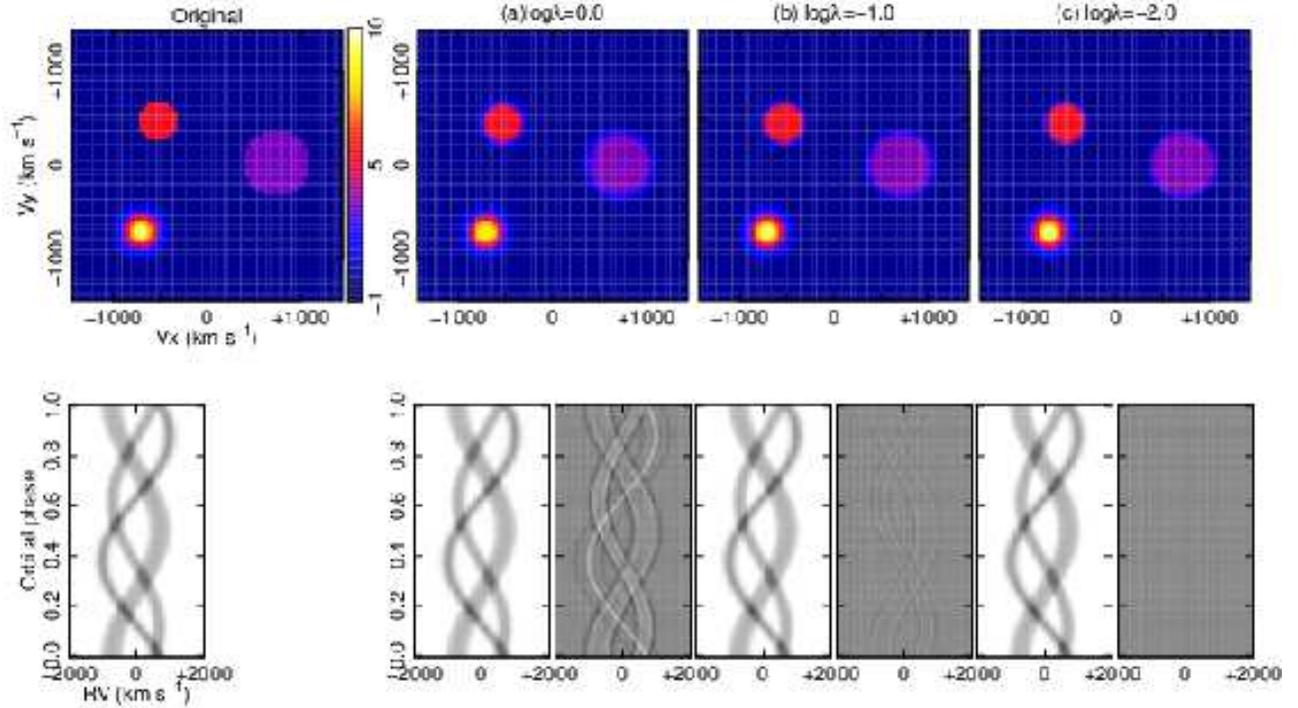}
  \end{center}
  \caption{Results of DTTVM using an artificial map having three
    spots. The artificial map is depicted in the upper left-most panel. 
    The simulated spectrum from the map is shown in the lower left-most
    panel. The spectra were calculated with a phase interval of 0.01
    and a radial velocity interval of $10\; {\rm km\,s^{-1}}$ between
    $-2000$ and $+2000\;{\rm km\, s^{-1}}$. Three sets of results are
    shown with different values of $\lambda$, which are indicated by the top of
    each panel. For each set, the estimated Doppler map is shown in
    the upper panel. The lower left and right panels are the model
    spectra and residuals between the original and model spectra,
    respectively. The Doppler maps consist of $64\times 64$ bins. The
    intensities of the assumed spots are 10, 5, and 3 at the peak of
    the lower left Gaussian profile and in the upper left and right
    flat-top regions, respectively. The intensity is 0 outside these
    spots.}\label{fig:dopmap_demof}  
\end{figure*}

We make artificial Doppler maps, then simulate line-profile 
variations, and reconstruct the maps from the simulated data. First,
we test DTTVM on a case in which the number of the data is larger than
the number of pixels of the Doppler map. Here, the number of the data
means the dimension of $\bm{d}$ in equation~(2), that is, the number
of all data points of the flux in all $v_{\rm R}$ and $\phi$. The
artificial map contains three spot structures, as shown in the upper
left-most panel of figure~\ref{fig:dopmap_demof}. Spectra were
calculated from this map with a phase interval of 0.01 and a radial
velocity interval of $10\;{\rm km\,s^{-1}}$ between $-2000$ and
$+2000\;{\rm km\,s^{-1}}$. No noise was added to the simulated
spectra. We assumed $\sigma$ in equation~(3) to be $100\; {\rm
km\,s^{-1}}$, which corresponded to $235\; {\rm km\,s^{-1}}$ in 
full-width of half maximum (FWHM) of the Gaussian profile. As a
result, the number of the input data is $40100$. Using this data,
DTTVM estimates $64\times64=4096$ elements of the Doppler
map. Figure~\ref{fig:dopmap_demof} shows the results. Three different
values of the hyperparameter, $\lambda$ were used to see the
dependence on it. Panels~(a) and (c) show the cases of the smallest
and largest $\lambda$. 

As can be seen in figure~\ref{fig:dopmap_demof}, the residuals
of the model from the assumed spectra are smaller for smaller
$\lambda$. This behavior can be easily understood because the solution
becomes close to that determined by the least-square term in the case
of very small $\lambda$ in equation~(2). No major difference can be
seen in all three estimated Doppler maps.

\begin{figure*}
  \begin{center}
    \FigureFile(170mm,170mm){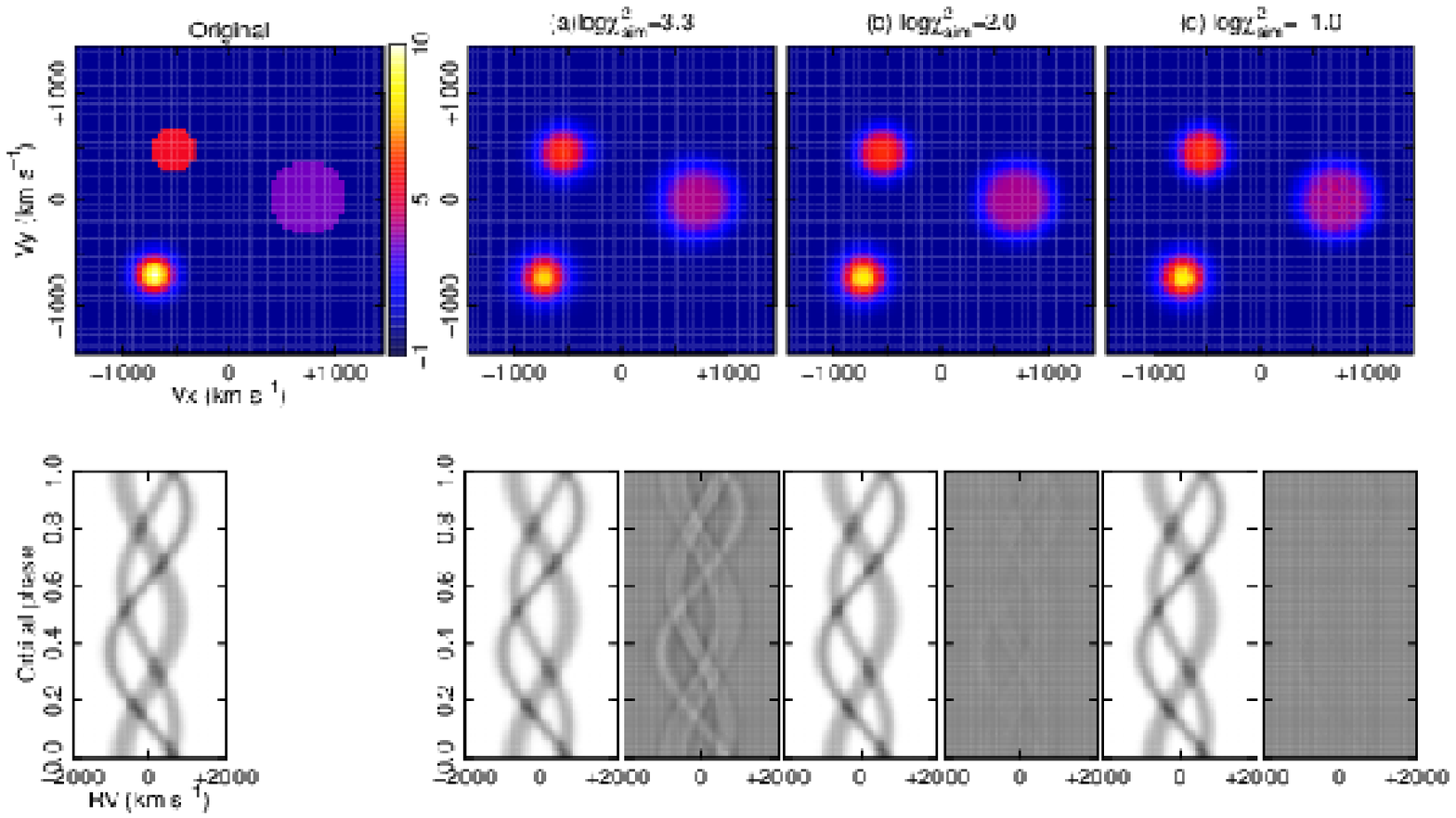}
  \end{center}
  \caption{As in figure~1, but for the MEM case. The ranges of
    the gray scale of the model and residual spectra are common in
    that in figure~1.}\label{fig:dopmap_demofMEM} 
\end{figure*} 

We also performed MEM reconstructions using the same map
as figure~1. We used two calculation codes: \texttt{doppler}
developed by T. Marsh\footnote{$\langle$http://deneb.astro.warwick.ac.uk/phsaap/software/doppler/html/OVERVIEW.html$\rangle$} and \texttt{dopp} developed by
H. Spruit
(\cite{spr98dt})\footnote{$\langle$http://www.mpa-garching.mpg.de/\~{}henk/$\rangle$}.
For the MEM models, the regularization term,
\begin{eqnarray}
\Phi_{\rm MEM}(\bm{s}) = -\sum_{i,j} \left( s_{i,j}
\ln{\frac{s_{i,j}}{\chi_{i,j}}} + \chi_{i,j} \right) 
\end{eqnarray} 
is used, where $\chi_{i,j}$ is the default image. According to 
\citet{spr98dt}, the standard default image is that created from
$\bm{s}$ by a Gaussian smearing, which is used in \texttt{dopp}. We
set the same default image also in \texttt{doppler}. We confirmed that
those two MEM codes provided almost consistent results (except for the
data of WZ~Sge, see subsection~3.2). All MEM results presented in this
paper are those obtained by \texttt{doppler}. Since we focus
on the reconstruction of sharp-edged profiles, we used a small
smearing factor in \texttt{doppler} (the parameter, BLURR$=1.0$). We
calculated the input spectra from the assumed maps by the method
defined in \texttt{doppler}.

Figure~\ref{fig:dopmap_demofMEM} shows the MEM results. In the code
\texttt{doppler}, the parameter $\chi^2_{\rm aim}$, or CAIM in the
code, corresponds to the hyper-parameter in the MEM model. As in
figure~3, we show three sets of results obtained by different
$\chi^2_{\rm aim}$, which are shown in the top of each panel. All
three results are consistent with those in the TVM case. We calculated
the RMS of the residuals between the model and data of spectra. In
cases~(c) of both figures~\ref{fig:dopmap_demof} and
\ref{fig:dopmap_demofMEM}, RMS are below 0.005, corresponding to 3~\%
of the peak flux in the spectra. Those two experiments demonstrate
that both TVM and MEM codes can reproduce the Doppler map when the
number of the data is large. 

\begin{figure*}
  \begin{center}
    \FigureFile(170mm,170mm){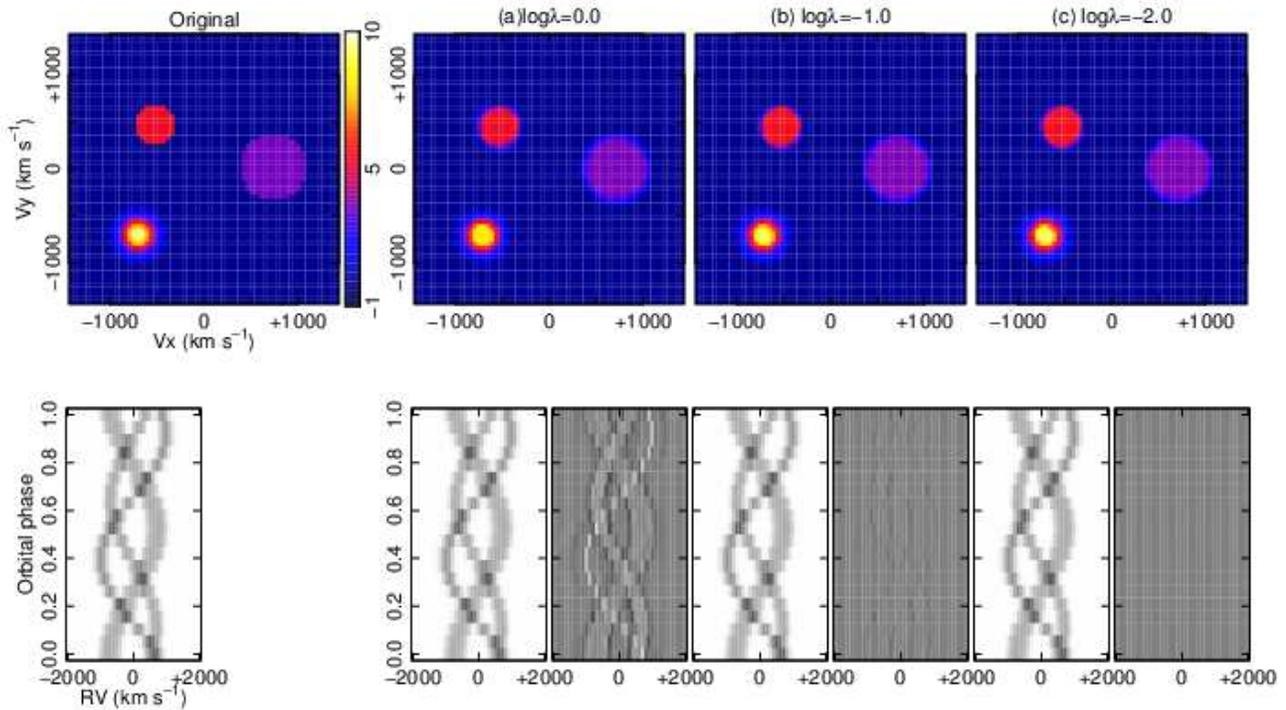}
  \end{center}
  \caption{As in figure~1, but for the small number of input data. The
    spectrum was calculated with a phase interval of 0.05 and a radial
    velocity interval of $100\; {\rm km\,s^{-1}}$ between $-2000$ and
    $+2000\;{\rm km\, s^{-1}}$.}\label{fig:dopmap_demo} 
\end{figure*}

Next, we test DTTVM in the case that the number of data is smaller
than the pixels of the Doppler map. Spectra were calculated from the
same map in figure~\ref{fig:dopmap_demof} with a phase interval of
0.05 and a radial velocity interval of $100\; {\rm km\,s^{-1}}$. The
instrument response is the same as the last case: $\sigma=100\; {\rm
  km\,s^{-1}}$, or ${\rm FWHM}=235\; {\rm km\,s^{-1}}$. The number of
the input data is $820$, which is less than the number of elements of
the estimated map, that is, $64\times64=4096$. The results are shown
in figure~\ref{fig:dopmap_demo}. DTTVM successfully reproduced the
original map in all three cases. In the assumed Doppler map, the upper
two spots have sharp-edge and flat-top profiles with different size
and intensity. The lower left spot has a Gaussian profile. This
difference in the structure is properly reproduced in the estimated
maps in figure~\ref{fig:dopmap_demo}. 

\begin{figure*}
  \begin{center}
    \FigureFile(170mm,170mm){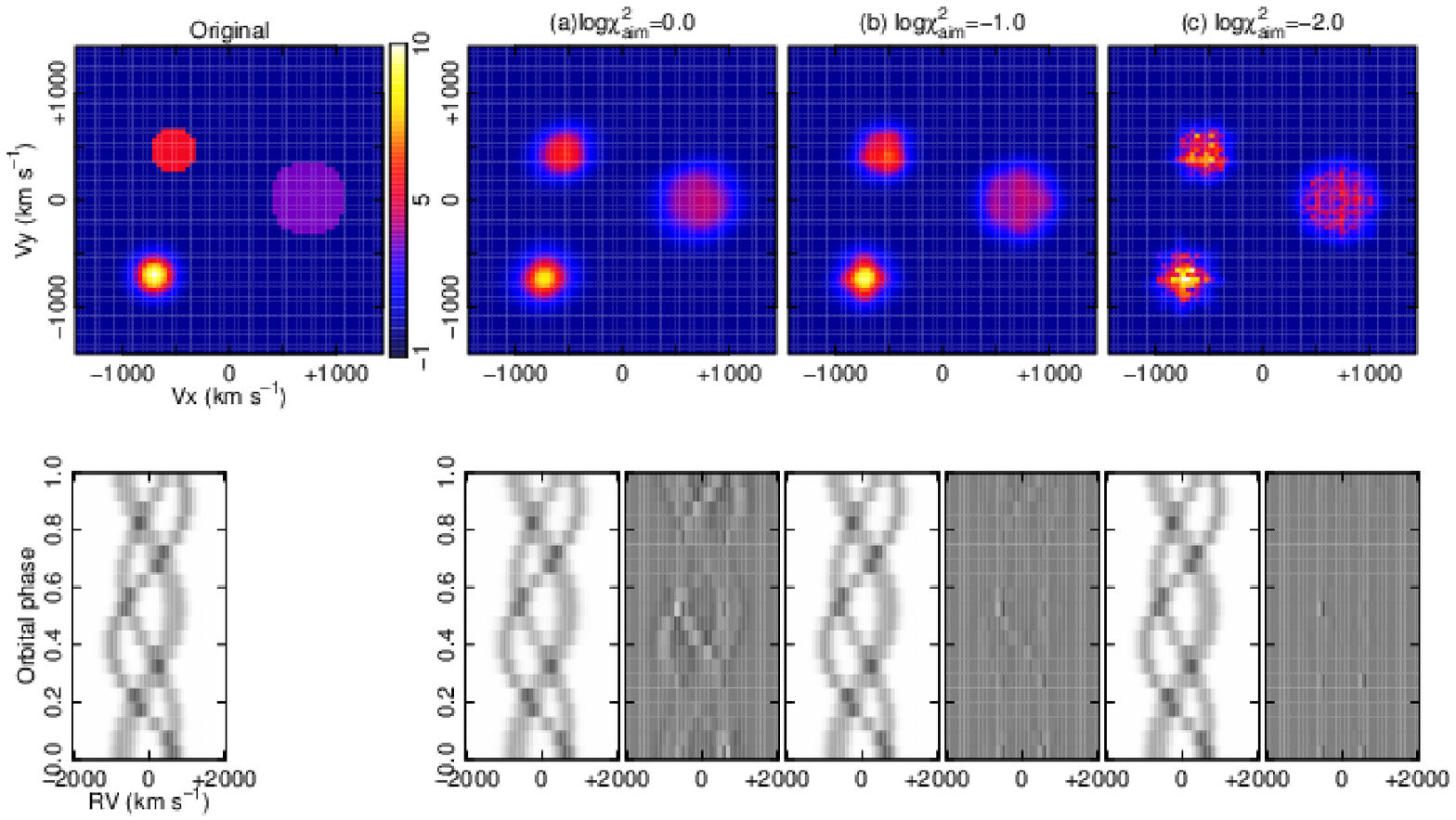}
  \end{center}
  \caption{As in figure~3, but for the MEM case. The color scales of
    the Doppler maps are different in each case, but the model spectra
    and residuals are shown in the same scale for each. The ranges of
    the gray scale of the model and residual spectra are common in
    that in figure~3.}\label{fig:dopmap_demoMEM} 
\end{figure*} 

Figure~\ref{fig:dopmap_demoMEM} is the result of MEM using the same
configuration of the data as in figure~\ref{fig:dopmap_demo}. The map of case~(a)
underestimates the flux of all spots. As a result, we can see 
significant residuals between the model and assumed spectra. Smaller
$\chi^2_{\rm aim}$ leads to smaller spectral residuals, while more spurious
structures appears in the Doppler maps, as shown in cases~(b) and
(c). The map is filled by noise in case~(c) and also cases with
further smaller $\chi^2_{\rm aim}$. 

The results shown in figures~1--4 suggest that DTTVM can
reconstruct both sharp-edge and smooth profiles in the Doppler map. This
characteristic is emphasized compared with the MEM method in the case
that the number of data is small.

\begin{figure*}
  \begin{center}
    \FigureFile(170mm,170mm){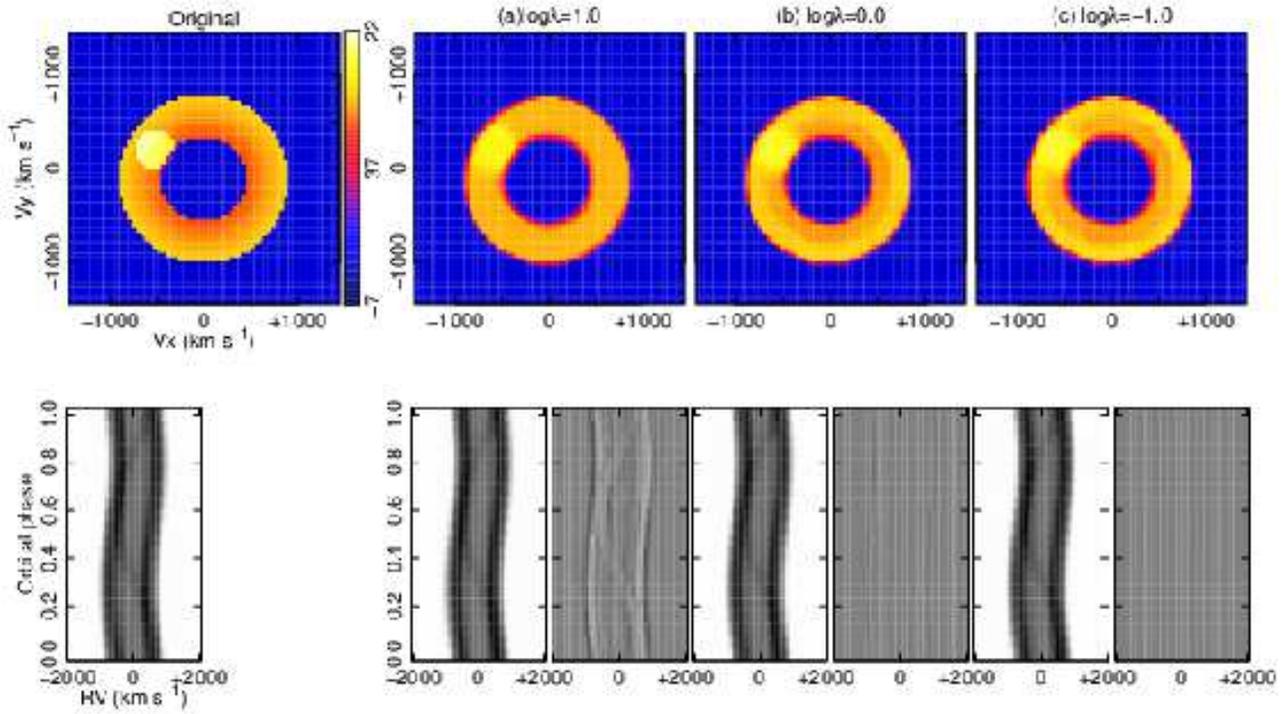}
  \end{center}
  \caption{As in figure~\ref{fig:dopmap_demo}, but for the artificial
    map having a disk and spot.}\label{fig:dopmap_demoDS}  
\end{figure*} 

We see another examples of the DTTVM reconstruction using artificial
maps in the case of the same small data set used in figure~3, in which a
phase interval is set to be 0.05 and a radial velocity interval is set
to be $100\; {\rm km\,s^{-1}}$. 
Figure~\ref{fig:dopmap_demoDS} shows the case of the Doppler map
having disk and spot features. In the accretion disks of CVs, the
line intensity is proposed to be a function of radius, $I \propto
r^{-3/2}$ (\cite{you81htcas}). In conjunction with the Kepler
rotation, $v \propto r^{-1/2}$, we set a small gradient for the disk
intensity, $I\propto r^{1/3}$ in those maps. The spot has a
constant intensity. The reconstructed maps show similar behavior as in
the previous case; the spot intensity is underestimated in a 
low $\lambda$ case, and the residuals of spectra are smaller in
higher $\lambda$ cases. The small intensity gradient is also
reconstructed in case~(b) and (c). 

\begin{figure*}
  \begin{center}
    \FigureFile(170mm,170mm){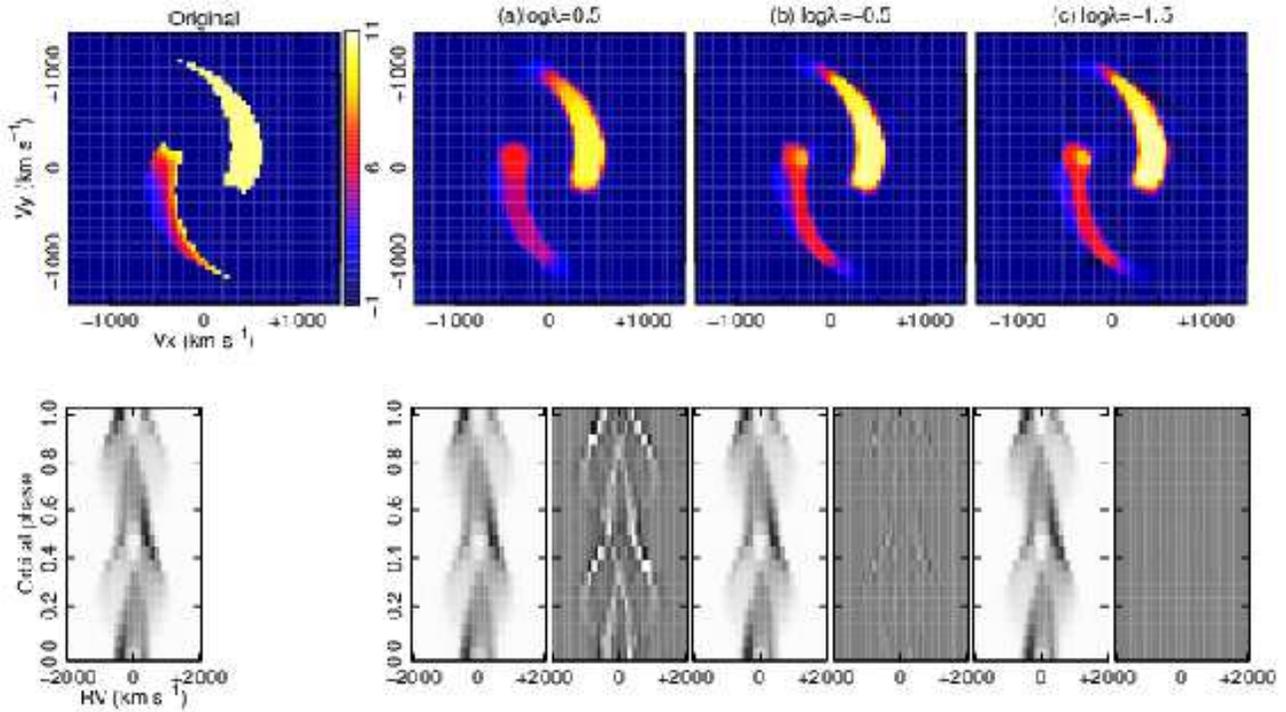}
  \end{center}
  \caption{As in figure~\ref{fig:dopmap_demo}, but for the artificial
    map having spiral patterns.}\label{fig:dopmap_demoSP}  
\end{figure*} 

Figure~\ref{fig:dopmap_demoSP} shows the case of the Doppler map having
two-armed spiral patterns. The upper spiral has a constant intensity,
and the lower has an intensity gradient with an exponential form. The
reconstructed maps by TVM again shows similar behavior that the
last cases. In this case, however, they fail to
reconstruct fine structures, such as a tail of the spirals and the
intensity gradient in the lower spiral, while sings of their features
are still seen in the maps. The low quality of those fine structures
is due to the finite spectral resolution of the simulated spectra. The
resolution was set to be $100\; {\rm km\,s^{-1}}$. The structure
smaller than this resolution lost its information in the simulated
spectra, as a result, it cannot be recovered. 

\subsection{Application for real data}

\begin{figure*}
  \begin{center}
    \FigureFile(150mm,170mm){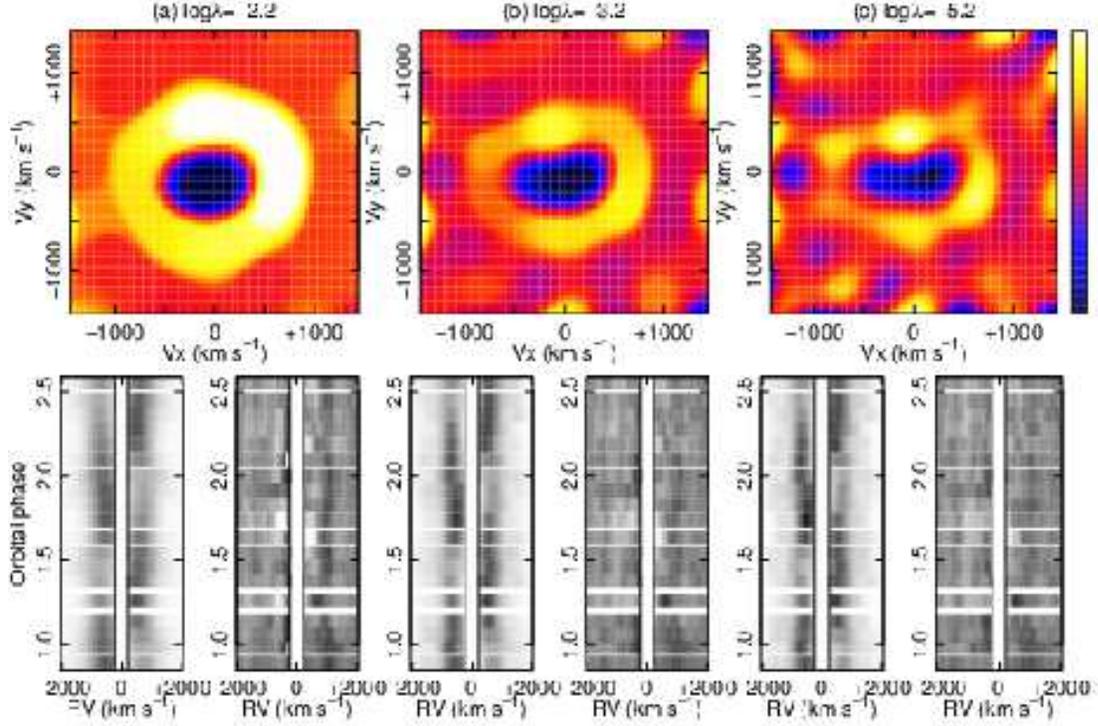}
  \end{center}
  \caption{Results of DTTVM for the data of WZ~Sge. Upper panels: the
    Doppler maps. Lower left and right panels: the model spectra and
    residuals between the model and observed spectra,
    respectively. The three sets of the panels were calculated with
    different $\lambda$, which are indicated in the figure. The
    Doppler maps consist of $64\times 64$ bins. The ranges of the gray
    scale of the model and residual spectra are common in that in
    figure~8. The data in the masked area in the spectra between
    $-200$ and $+200\;{\rm km\, s^{-1}}$ was not used for the Doppler
    tomography because the contamination of the absorption core is
    strong.}\label{fig:dopmap_wzsge} 
\end{figure*}

We apply DTTVM to the data of the dwarf nova, WZ~Sge taken on the 10th 
day of the superoutburst in 2001 (\cite{nog04wzsge}). The number
of the data point is 630. We estimated a $64\times 64$ Doppler map
using the data. The data having radial velocities between $-200$ and
$+200\;{\rm km\, s^{-1}}$ were excluded because of the prominent
absorption core which violates the model assumption of the Doppler
tomography. The velocity resolution was $270\;{\rm km\, s^{-1}}$, 
which was used as $\sigma$ in
equation~(3). Figure~\ref{fig:dopmap_wzsge} presents the Doppler maps
and model spectra. The results were computed using 
$\log_{10}\lambda$ of (a) $-2.2$, (b) $-3.2$, and (c) $-5.2$. All
Doppler maps exhibit a disk feature having non-axisymmetric intensity
distribution. Its entire structure is circular in case~(a), but
elliptical in cases~(b) and (c). The model spectra reproduce the
observed one with a high precision in cases~(b) and (c), while we can
see several significant components of residuals in case~(a).

\begin{figure*}
  \begin{center}
    \FigureFile(150mm,170mm){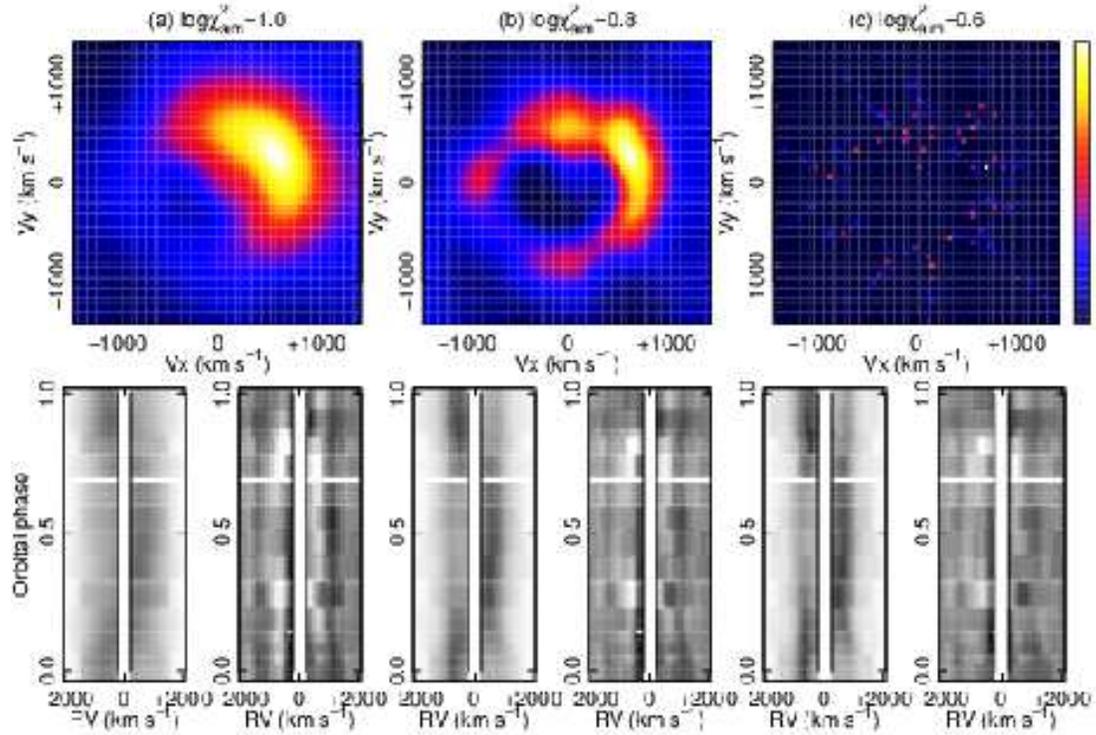}
  \end{center}
  \caption{As in figure~\ref{fig:dopmap_wzsge}, but for the MEM
    case. The color scales of the Doppler maps are different in each
    case. The ranges of the gray scale of the model and
    residual spectra are common in that in figure~7. The
    orbital phase in the trailed spectra is set from 0.0 to 1.0, not
    as in figure~\ref{fig:dopmap_wzsge}. This is a feature of the
    calculation code, and the spectra were not binned in the phase for
  the analysis.}\label{fig:dopmap_wzsgeMEM} 
\end{figure*}

Figure~\ref{fig:dopmap_wzsgeMEM} is the result of the Doppler
tomography for the same data calculated using MEM. In case~(a), the
bright region appears around the upper-right area in the Doppler map,
while the map is so simple that the spectral residuals of the model
from the data is large. Reducing $\chi^2_{\rm aim}$ leads to reducing
the residuals, as can be seen in case~(b). In addition to the
upper-right region, the map has another bright region at
$(v_x,v_y)\sim(0,500)$ and a weak sign of a circular
disk. \citet{ste04doppler} performed Doppler tomography with MEM also
using data of WZ~Sge in outburst. Their H$\beta$ map on the ninth day
of the superoutburst has similar features to case~(b). In our MEM
analysis, noisy structures appear in case~(c) and calculations with
further smaller $\chi^2_{\rm aim}$, but they are evidently
artifacts. In cases~(a) and (b), there are significant residuals of
the spectral data at $+500\; {\rm km \, s^{-1}}$ for phase
$0.75$. This feature has counterparts at $-500\; {\rm km \, s^{-1}}$
for phase $0.25$. Another local residuals can be seen at $-500\; {\rm
  km \, s^{-1}}$ for phase $0.8$. 

Table~1 lists the RMS of residuals between the model and observed
spectra of WZ~Sge for both cases of TVM and MEM. Note that the spectra
are normalized by the continuum level, and the average and maximum
fluxes are 0.06 and 0.26, respectively. As shown in this
table, the RMS obtained by DTTVM are smaller than those by MEM.
A reason for the high quality reproduction of DTTVM may be the
constraint condition of $\bm{s}$ in equation~(2): the elements of
$\bm{s}$ can be negative in our model, but must be positive in the
available MEM codes. The line profile observed in CVs frequently has
an absorption core. This absorption is formed by the outermost, low
radial-velocity region of the disk where the line emission is
partially optically-thick. Our model can fit the absorption
core by taking negative values of $\bm{s}$, which is prohibited in the
MEM codes. The low velocity regions between $-200$ and $+200\;{\rm
km\, s^{-1}}$ are masked for the current analysis, while it may not be
enough to mask the entire absorption component.

\begin{table*}
\caption{RMS of the residuals between the model and observed spectra
  of WZ~Sge.}\label{tab:rms}
\begin{center}
\begin{tabular}{cclll}
  \hline
  Model & Figure number & \multicolumn{3}{c}{RMS}\\ 
  & & case (a) & case (b) & case (c) \\
  \hline
  TVM & \ref{fig:dopmap_wzsge} & 0.023 & 0.019 & 0.017 \\
  MEM & \ref{fig:dopmap_wzsgeMEM} & 0.031 & 0.024 & 0.020 \\
  TVM (absorption core corrected) & \ref{fig:dopmap_gadd} & 0.022 & 0.019 & 0.017 \\
  MEM (absorption core corrected) & \ref{fig:dopmapMEM_gadd} & 0.022 & 0.018 & 0.016 \\ 
  \hline
\end{tabular}
\end{center}
\end{table*}

\begin{figure*}
  \begin{center}
    \FigureFile(150mm,170mm){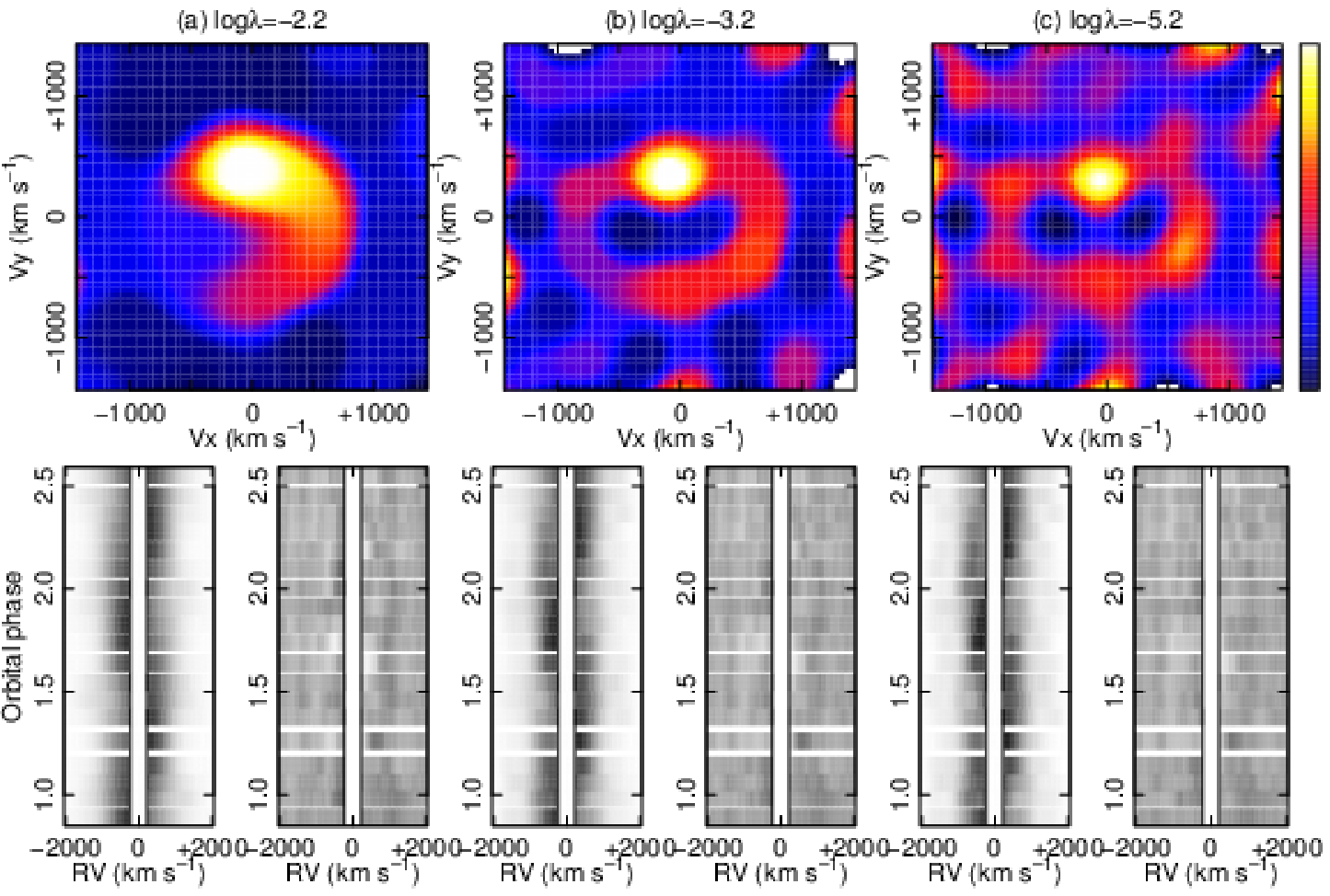}
  \end{center}
  \caption{As in figure~7, but for the data of WZ~Sge whose
      absorption core is corrected by adding a Gaussian
      component.}\label{fig:dopmap_gadd} 
\end{figure*}

\begin{figure*}
  \begin{center}
    \FigureFile(150mm,170mm){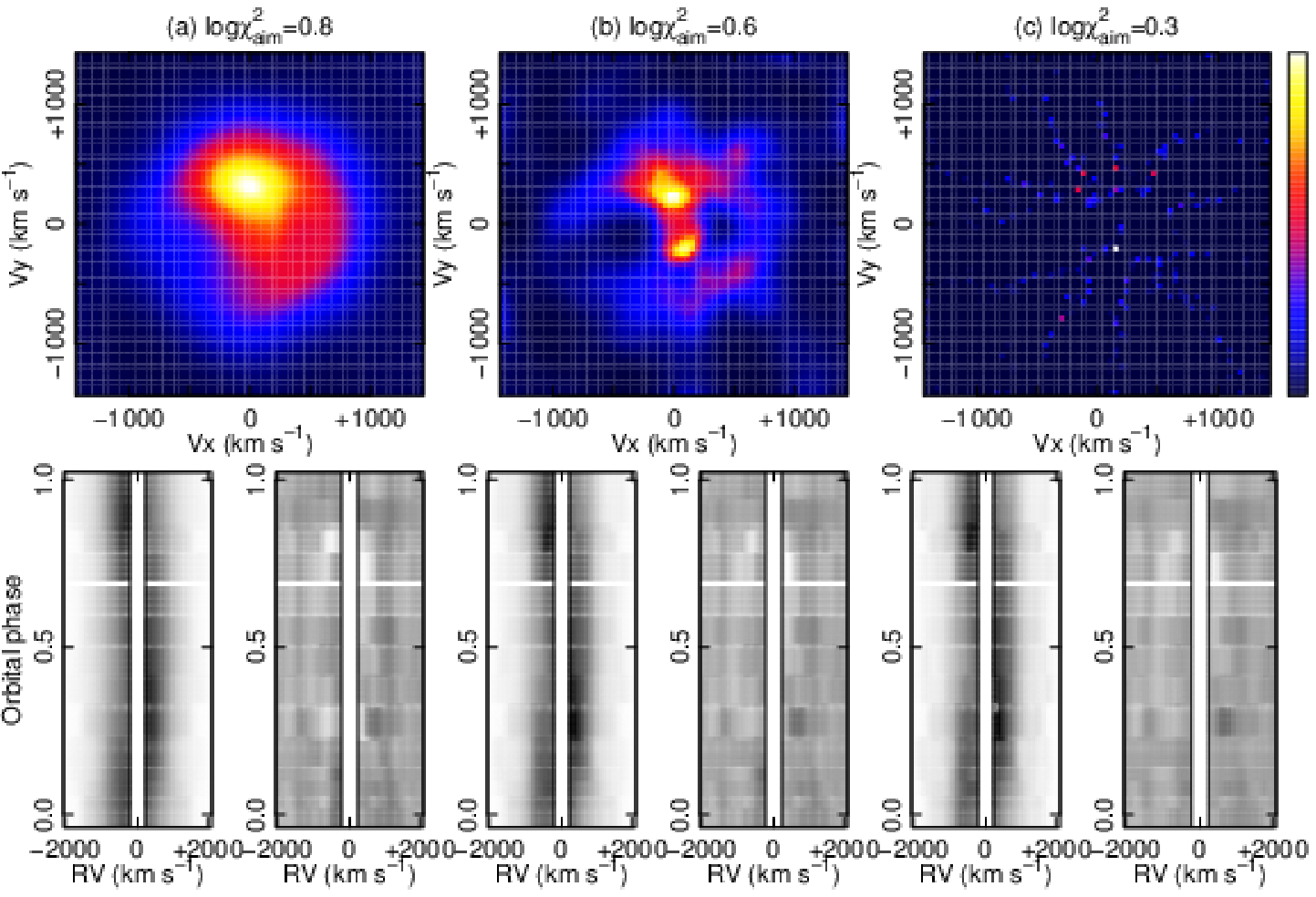}
  \end{center}
  \caption{As in figure~\ref{fig:dopmap_gadd}, but for the MEM
      case.}\label{fig:dopmapMEM_gadd} 
\end{figure*}

Then, we added a Gaussian component to the data so that the negative
pixels at the low velocity area in the Doppler map disappears
(\cite{mar90ugem}; \cite{har96oycar}). The Gaussian component has a
standard deviation of $300\;{\rm km\,s^{-1}}$ and the peak
(normalized) flux of 0.25, centering at a radial velocity of 
$0.0\;{\rm km\,s^{-1}}$. These are minimum values by which the central
negative pixels disappear in the Doppler map estimated by
DTTVM. We analyzed the modified data using DTTVM and the MEM code. The
results are shown in figures~\ref{fig:dopmap_gadd} and
\ref{fig:dopmapMEM_gadd}, respectively. 

The DTTVM results (figure~\ref{fig:dopmap_gadd}) are
consistent with those in figure~\ref{fig:dopmap_wzsge}: The
brightest spot lies at $(v_x,v_y)\sim(0,+500)$, and the lower-right
area in the disk region is also bright in the Doppler maps. There
are several high-velocity areas having very low intensity, or negative
values in some pixels in figure~\ref{fig:dopmap_gadd}. They are
probably due to over-fitting of the noise in the data. They are so
localized that they may make narrow absorption-like effect on the
spectra, but do not change the entire structure of the Doppler map. 

In the case of the MEM (figure~\ref{fig:dopmapMEM_gadd}),
especially in case~(b), the Doppler maps have a different feature
from those in figure~\ref{fig:dopmap_wzsgeMEM}. In addition to the
brightest spot at $(v_x,v_y)\sim(0,+500)$, we can see a weak sign of
the bright region  in the lower-right area of the map. This feature is
rather similar to the DTTVM results. In addition, the RMS of the model
spectra to the data is small comparable to the DTTVM values, as shown
in table~1. We note that these behavior was only seen in the code
\texttt{doppler}, and not in \texttt{dopp}. In the case of
\texttt{dopp}, the Doppler maps little depend on the presence or
absence of the absorption component, and the upper-right disk area is
always the brightest, as in figure~\ref{fig:dopmap_wzsgeMEM}.

\begin{figure*}
  \begin{center}
    \FigureFile(150mm,170mm){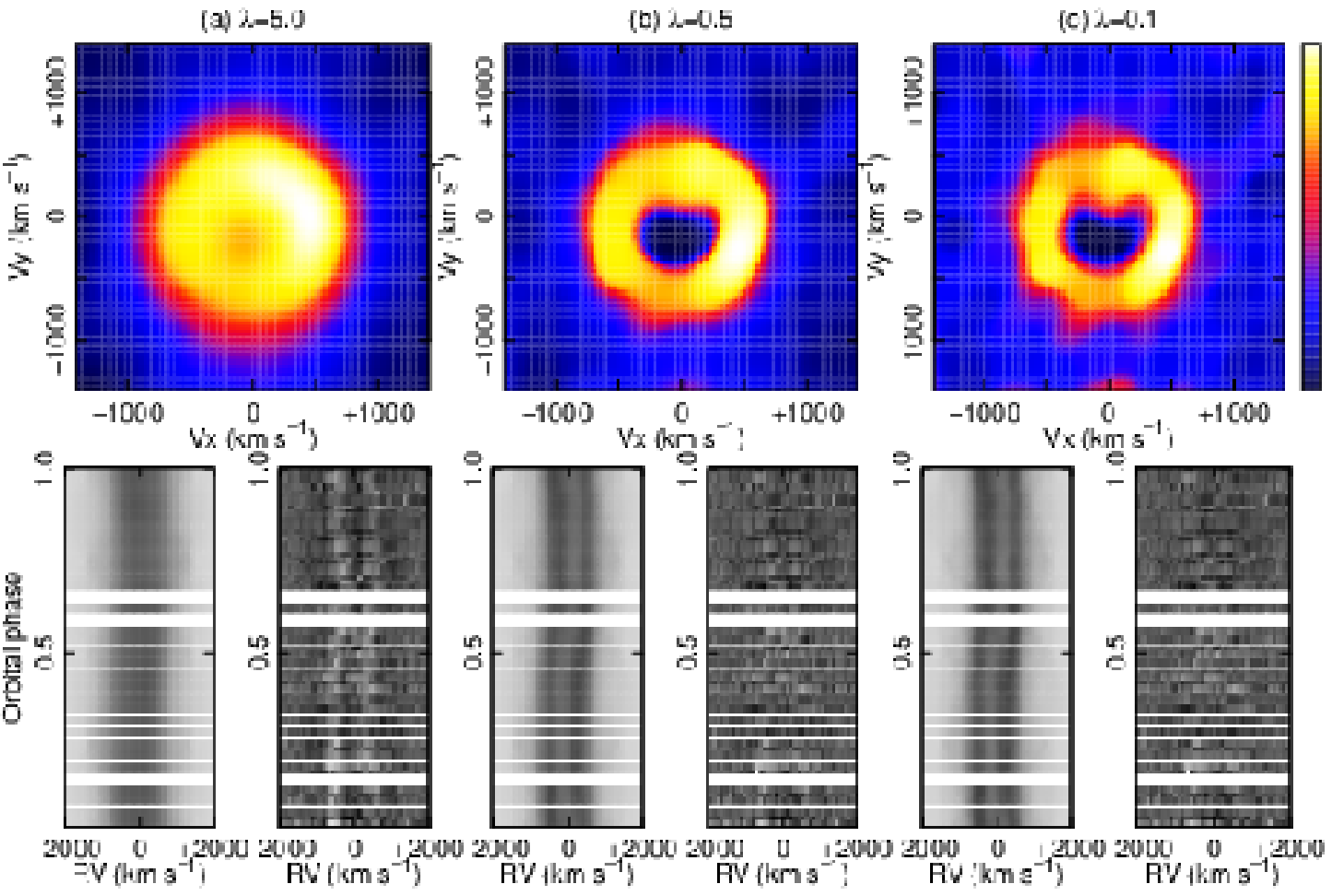}
  \end{center}
  \caption{As in figure~\ref{fig:dopmap_wzsge}, but for the data of
    TU~Men.}\label{fig:dopmap_tumen} 
\end{figure*}

We also analyzed the data of a dwarf nova TU~Men in quiescence. The
data was reported in \citet{men95tumen}. The total number of the data
point is 5293. We estimated the $80\times80$ Doppler
map. Figure~\ref{fig:dopmap_tumen} shows the result obtained by
DTTVM. The results were computed using $\lambda$ of (a)
$5.0$, (b) $0.5$, and (c) $0.1$. In cases~(b) and (c), a sign of the 
secondary star can be seen at $(v_x,v_y)\sim(0,100)$ in the Doppler
map, as well as a disk structure. \citet{tap03dopmap} also report the
result of the Doppler tomography with MEM using the same data. Their
result is consistent with figure~\ref{fig:dopmap_tumen}, except for
spectral residuals. Their result shows several rotating components in
the residuals. They propose that the residuals are due to the
anisotropy of the emitting region. However, our result exhibits no
major residual, indicating that the MEM method failed to recover
localized components. 

\subsection{Cross-validation method to determine a hyperparameter,
  $\lambda$} 

DTTVM has a hyperparameter, $\lambda$, as shown in equation~(2). The
reasonable $\lambda$ can be estimated by the data itself using the
cross-validation method. In the $K$-fold cross-validation, the 
data is divided into $K$ roughly equal parts, $\bm{y}_k$
($k=1$--$K$). For each $k$, the training data is defined as the all
$K-1$ data except for the validation data, $\bm{y}_k$. The
optimization of the model using the training data gives
$\hat{\bm{x}}_{k,\lambda}$ at a certain $\lambda$. The prediction
reliability of the model is evaluated with mean square errors (MSE);
\begin{eqnarray}
E(\lambda) = \frac{1}{K} \sum_{k=1}^K E_k(\lambda)\\
E_k(\lambda) = \frac{1}{N_k} \| \bm{y}_k - A\hat{\bm{x}}_{k,\lambda} \|^2_2,
\end{eqnarray}
where $N_k$ is the number of the validation data, $\bm{y}_k$.
The MSE, $E(\lambda)$ can be the minimum value at a
certain $\lambda$. In a large $\lambda$ regime, the least-square
term can be large, which leads to a large $E(\lambda)$. In a small
$\lambda$ regime, on the other hand, the model can reproduce even
noises in the training data, and thereby have a low prediction
reliability, and lead to a large $E(\lambda)$. This situation is known
as model over-fitting. Thus, we can determine a reasonable range of
$\lambda$ by the behavior of $E(\lambda)$. We set $K=10$ in this paper.

\begin{figure}
  \begin{center}
    \FigureFile(75mm,170mm){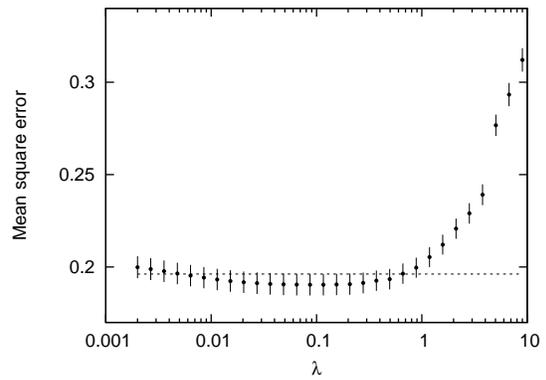}
  \end{center}
  \caption{The mean square errors between the model and validation
    data against a hyperparameter, $\lambda$ for the data of
    TU~Men.}\label{fig:cv_tumen}
\end{figure}

Figure~\ref{fig:cv_tumen} shows $E(\lambda)$ against $\lambda$,
calculated using the data of TU~Men. $E(\lambda)$ clearly takes the
minimum at $\lambda\sim0.1$. The vertical bars associated with
$E(\lambda)$ represent the standard error of $E_k(\lambda)$
($k=1$--$10$). Hence, there is an uncertainty of $\lambda$ at which
$E(\lambda)$ takes minimum, in other words, for the best
model. Instead of the model with the minimal $E(\lambda)$, we can take
the simplest model whose $E(\lambda)$ is within one standard error of
the minimal $E(\lambda)$. This is sometimes called as the ``one
standard error rule''. The three Doppler maps shown in
figure~\ref{fig:dopmap_tumen} corresponds to the results for 
$\lambda=5.0$, $0.5$, and $0.1$. The model with $\lambda=5.0$ is so
simple, or sparse in the gradient domain that $E(\lambda)$ is
large. The models with $\lambda=0.5$ and $0.1$ correspond to those
obtained by the one standard error rule and minimal $E(\lambda)$,
respectively. We can consider that the case for $\lambda=0.5$ is the
most reasonable model for the data. 

\begin{figure}
  \begin{center}
    \FigureFile(75mm,170mm){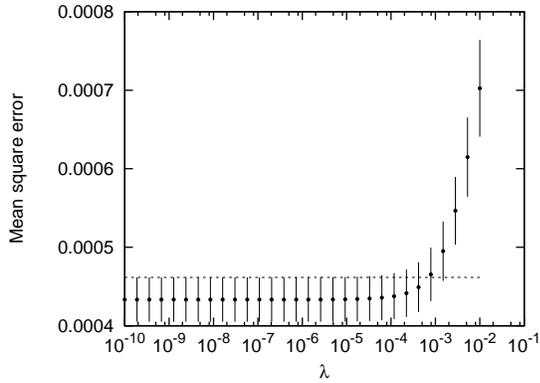}
  \end{center}
  \caption{The same as figure~\ref{fig:cv_tumen}, but for the data of
    WZ~Sge. The dashed line indicates the MSE at
    $\lambda=10^{-7}$.}\label{fig:cv_wzsge}
\end{figure}

Figure~\ref{fig:cv_wzsge} shows the same as figure~\ref{fig:cv_tumen},
but for the data of WZ~Sge. In this case, MSE takes the minimum at
$\log_{10} \lambda=-5.2$, while the increasing trend of $E(\lambda)$ in
small $\lambda$ is not prominent compared with the case of TU~Men. In
figure~\ref{fig:dopmap_wzsge}, we show the results for $\log_{10}
\lambda=-2.2$, $-3.2$, and $-5.2$, that is, under-fitted model, model
with $\lambda$ for the one standard error rule, and model with the
minimal $E(\lambda)$. As in the case of TU~Men, we can consider that
panel~(b) of figure~4 is the most reasonable model.

\section{Discussion} 

As demonstrated in subsection~3.1, DTTVM can reconstruct both
sharp-edge and smooth profiles in the Doppler maps even in the case
that the number of data is much smaller than the number of pixels of
the map. Hence, this model is advantageous for the reconstruction of
highly localized or edge structures in the Doppler map, such as the
emission from the secondary star, hot spot, or shock wavefront. 

Applications to the real
data of WZ~Sge and TU~Men suggest that DTTVM can reconstruct the
observed spectra with a high accuracy. As shown in subsection~3.2, we
confirmed that DTTVM can provide the Doppler maps
which are consistent with those by MEM. On the other hand, compared
with the TVM maps, the MEM maps were more sensitive to the values of
hyper-parameter: The maps tend to be very noisy with the
hyper-parameter ($\chi^2_{\rm aim}$ in \texttt{doppler}) below a
threshold. This is a common feature in \texttt{doppler} and
\texttt{dopp}. The correction for the absorption core 
can also change the feature of the maps, as shown in the case of the
data of WZ~Sge (figures~\ref{fig:dopmap_wzsgeMEM} and
\ref{fig:dopmapMEM_gadd}). The dependency on those factors was
relatively low in DTTVM (figures~\ref{fig:dopmap_wzsge} and 
\ref{fig:dopmap_gadd}). In conjunction with the cross-validation
method, DTTVM can provide robust estimates of the Doppler map,
compared with the MEM methods, especially in the case of that the
number of data is small.

\begin{figure}
  \begin{center}
    \FigureFile(75mm,170mm){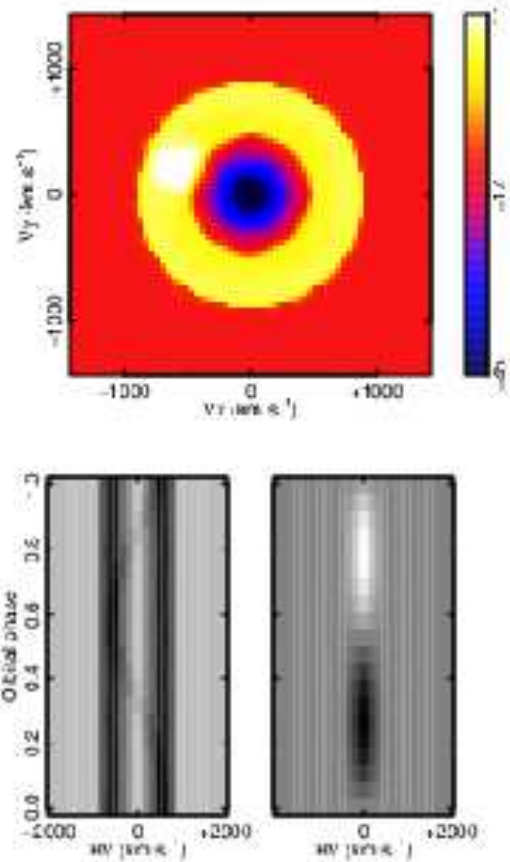}
  \end{center}
  \caption{The upper panel: Doppler map estimated from the spectra
    with an absorption core. The lower panels: the model (left) and
    residuals (right) of spectra. The assumed map is the same as
    figure~\ref{fig:dopmap_demoDS}.}\label{fig:dopmap_abs} 
\end{figure}

DTTVM can fit the absorption core by taking negative values of
$\bm{s}$. However, it breaks the assumption of the Doppler tomography
that all line sources rotate with the orbital motion of the binary. It
is not evident whether the Doppler map can be deformed by the presence
of absorption core. We performed a simple experiment using the
artificial data. We assumed a Doppler map having a disk and spot,
which is similar to figure~\ref{fig:dopmap_demoDS}, but the velocity
of the disk center is set to be $(v_x,v_y)=(0,0)$ in this case for
simplicity. The observed trailed spectra were simulated from the
map. Then, we added the absorption core to the spectra, and estimated
the map by DTTVM. We assumed a periodic variation in the depth of the
absorption component, in other words, the amplitude was proportional to
$\sin(\phi)$. The amplitude variation was assumed in order to see how
DTTVM behaves when the strength of the absorption core depended on the
phase. Figure~\ref{fig:dopmap_abs} shows the estimated map, model
and residuals of spectra. The negative region appears at the center of
the map. Compared with figure~\ref{fig:dopmap_demoDS}, the general
features of the assumed disk and spot is unchanged by the presence of
the absorption core. The spectral residuals clearly indicate the
assume variation of the absorption component. The result of this
experiment suggests that the model reconstructed the average
absorption profile without significant change in the disk and spot
features.

\section{Summary}

We have developed a new model of the Doppler tomography using total
variation minimization (TVM). This method can reconstruct localized
features possibly having sharp-edges in the Doppler map, such as the 
emission from the secondary star, hot spot, and shocked region. This
characteristics are emphasized in the case that the number of the data
is much smaller than the number of pixels in the Doppler map. We
applied it to the data of WZ~Sge and TU~Men. The model reproduces
the observed spectra with a high precision. We demonstrated that
reasonable values of the hyperparameter in our model can be estimated
by the data itself using cross-validation.

\bigskip
This work was supported by JSPS KAKENHI Grant Number 22540252 and
25120007. We appreciate comments and suggestions from Dr. Shiro
Ikeda and anonymous referee. 
REM acknowledges support by Fondecyt grant 1110347 and the BASAL
Centro de Astrofisica y Tecnologias Afines (CATA) PFB--06/2007.


\end{document}